\documentclass[aps,10pt]{revtex4}
\usepackage{epsfig}

\begin{document}
\title{Finite-Difference Lattice Boltzmann Methods for Binary Fluids}
\author{Aiguo Xu}
\affiliation{Devision of Physics and Astronomy, Yoshida-South Campus, 
Kyoto University, 
Sakyo-ku, Kyoto, 606-8501, Japan }
\begin{abstract}
We investigate two-fluid BGK kinetic methods for binary fluids. 
The developed theory works for asymmetric as well as symmetric 
systems. For symmetric systems it recovers Sirovich's theory 
and is summarized in models A and B. 
For asymmetric systems it contributes models C, D and E which 
are especially useful when the total masses and/or local 
temperatures of the two components are greatly different. 
The kinetic models are discretized based on an octagonal 
discrete velocity model.
The discrete-velocity kinetic models and the continuous ones are 
required to describe the same hydrodynamic equations. 
The combination of a discrete-velocity kinetic model and an 
appropriate finite-difference scheme composes a finite-difference 
lattice Boltzmann method. 
The validity of the formulated methods is verified by
investigating (i) uniform relaxation processes, (ii) isothermal 
Couette flow, and (iii) diffusion behavior.

\end{abstract}

\maketitle

\section{Introduction}

Gas kinetic theory plays a fundamental role in understanding many complex
processes. To make solutions possible, many of the kinetic models for gases
are based on the linearized Boltzmann equation, especially based on the BGK
approximation\cite{PR94511}. Even thus, only in very limited cases are 
analytic solutions available. 
Basically speaking, there are two options to
simulate Boltzmann equation systems. First, one can design procedures based
on the fundamental properties of rarefied gas alone, like free flow, the
mean free path, and collision frequency. Such a scheme does not need an {\it %
a priori} relationship with the Boltzmann equation, but the scheme itself will
reflect many ideas and/or concepts used in the derivation of Boltzmann
equation. In the best case, such a simulation will produce results being 
consistent with the solution of Boltzmann equation. The second option is to
start from the Boltzmann equation and design numerical schemes as accurate
as possible\cite{Carlo}. 
The discrete Boltzmann equation approach or lattice
Boltzmann method (LBM) has been becoming a viable and promising 
scheme for simulating fluid flows\cite{Succi}.

LBMs for single-component fluids have been well studied, while for
binary mixtures still need more clarification\cite{Succip1}. For binary
fluids, although various LBMs have been proposed
\cite{PRE6635301,IJCES373,Yeomans2fm,PRE6835302,PRE6756105,PRA434320,PRE474247,PFA52557,JSP81379,EPL32463,PRL83576,ICCS2003,CEJP2382,VictorIJMPC,VictorPRE,ShanChen,Coveney,XuEPL}
, most of them \cite
{PRE6835302,PRE6756105,PRA434320,PRE474247,PFA52557,JSP81379,EPL32463,PRL83576,ICCS2003,CEJP2382,VictorIJMPC,VictorPRE}
are based on the single-fluid theory\cite{PhysA299494}. For systems
with different component properties, a two-fluid theory is necessary. 
Sirovich's two-fluid kinetic theory\cite{PF5908} works for (approximately)
symmetric systems where the two components have (approximately) the same 
total masses and local temperatures. A LBM based on Sirovich's theory
and for the complete two-dimensional Navier-Stokes equations(NSE) is given in 
\cite{XuEPL}. This LBM is based on a two-dimenaional
model with sixty-one discrete velocities (D2V61).  
Many compressible fluids can be well described by the Euler
equations\cite{Dalton}. In fluid mechamics of low-speed flow, the 
temperature remains nearly constant and consequently the isothermal 
NSE description is extensively used\cite{Dalton}. 
From the Chapman-Enskog procedure\cite{Chap} the Euler equation  is a
lower-order  approximation compared with the NSE. The isothermal NSE is a
simplified case of the complete NSE. 
For the above two kinds of systems, using the LBM for complete NSE 
system is not neccessary and computationally inefficient.
In this study we generalize Sirovich's theory so that it works also for 
asymmetric systems where the total masses and/or local temperatures of 
the two components are greatly different, then  
formulate LBMs for the two kinds of systems. 
The LBMs formulated here require simpler discrete velocity
models(DVMs). 
For the Euler-equation
system a DVM with thirty-three discrete velocities (D2V33) is
enough. For the isothermal NSE system, a D2V25 is sufficient.
 
This paper is arranged in the following way: In section II we review and
develop the two-fluid BGK kinetic theory. Sirovich's original
treatments are clarified and summarized in models A and B. 
For asymmetric systems three kinetic models (C, D and E) are derived. 
The hydrodynamics and diffusion behavior of the model systems are
discussed.
In section III the kinetic models
are discretized based on a multispeed discrete velocity model. 
Then, possible FD schemes are given and the corresponding numerical
viscosities and diffusivities are analyzed. Numerical tests are shown in
Section IV. Section V concludes the present paper.


\section{Two-fluid BGK kinetic theory}

In a binary system with two components, $A$ and $B$, roughly speaking, the
approach to equilibrium can be divided into two processes. One is
referred to as Maxwellization (i.e., each species equilibrates within itself 
so that the local distribution function approaches to its local
Maxwellian). The other is the equilibration of species (i.e., 
the differences in hydrodynamic velocities and local temperatures of the 
two components eventually vanish). 
Correspondingly, the interparticle collisions fall into two categories:
self-collisions (collisions within the same species) and cross-collisions
(collisions between different species)\cite{PF5908,PRE6635301}.

\subsection{General description}

For a two-dimensional binary gas system the BGK kinetic equations
read\cite{PRE6635301}, 
\begin{equation}
\partial _tf^A+{\bf v}^A{\bf \cdot }\frac \partial {\partial {\bf r}}f^A+%
{\bf a}^A\cdot \frac \partial {\partial {\bf v}^A}f^A=J^{AA}+J^{AB}\text{,}
\label{ce1}
\end{equation}
\begin{equation}
\partial _tf^B+{\bf v}^B{\bf \cdot }\frac \partial {\partial {\bf r}}f^B+%
{\bf a}^B\cdot \frac \partial {\partial {\bf v}^B}f^B=J^{BB}+J^{BA}\text{,}
\label{ce1b}
\end{equation}
where 
\begin{equation}
J^{AA}=-\frac 1{\tau ^{AA}}\left[ f^A-f^{A(0)}\right] \text{, }J^{AB}=-\frac %
1{\tau ^{AB}}\left[ f^A-f^{AB(0)}\right] \text{,}  \label{ce2}
\end{equation}
\begin{equation}
f^{A(0)}=\frac{n^A}{2\pi \Theta ^A}\exp \left[ -\frac{\left( {\bf v}^A-{\bf u%
}^A\right) ^2}{2\Theta ^A}\right] \text{,}  \label{ce3}
\end{equation}
\begin{equation}
f^{AB(0)}=\frac{n^A}{2\pi \Theta ^{AB}}\exp \left[ -\frac{\left( {\bf v}^A-%
{\bf u}\right) ^2}{2\Theta ^{AB}}\right] \text{,}  \label{ce4}
\end{equation}
\begin{equation}
\Theta ^A=\frac{k_BT^A}{m^A}\text{, }\Theta ^{AB}=\frac{k_BT}{m^A}\text{.}
\label{ce4a}
\end{equation}
$f^A$ ( $f^B$) and ${\bf v}^A$ (${\bf v}^B$) are the distribution function
and particle velocity of the component $A$ ($B$); $f^{A(0)}$ and $f^{AB(0)%
\text{ }}$are the local Maxwellians which work as references for the self-
and cross-collisions; $n^A$, ${\bf u}^A$, $T^A$ are the local number density,
hydrodynamic velocity and temperature of the species $A$; ${\bf u}$, $%
T $ are the local hydrodynamic velocity and temperature of the mixture after
equilibration process; ${\bf a}^A$ is the acceleration of the species $A$
due to the effective external field. For
species $A$, we have 
\begin{equation}
n^A=\int d{\bf v}^Af^A\text{, }  \label{ce5}
\end{equation}
\begin{equation}
n^A{\bf u}^A=\int d{\bf v}^A{\bf v}^Af^A\text{,}  \label{ce6}
\end{equation}
\begin{equation}
n^Ak_BT^A=\int d{\bf v}^A\frac 12m^A\left( {\bf v}^A-{\bf u}^A\right) ^2f^A%
\text{,}  \label{ce7}
\end{equation}
\begin{equation}
\rho ^A=n^Am^A\text{,}  \label{ce8}
\end{equation}
\begin{equation}
e_{\text{int}}^A\text{ (}P_0^A\text{)}=n^Ak_BT^A\text{,}  \label{ce8a}
\end{equation}
where $\rho^A$ and $e_{\text{int}}^A$ ($P_0^A$) are the local mass
density and internal mean kinetic energy
(hydrostatic pressure) of species $A$, $k_B$ is the Boltzmann constant.
 For species $B$, we have similar
relations.

For the mixture, we have 
\begin{equation}
n=n^A+n^B\text{, }\rho =\rho ^A+\rho ^B\text{,}  \label{ce9}
\end{equation}
\begin{equation}
{\bf u}=\frac{\rho ^A{\bf u}^A+\rho ^B{\bf u}^B}\rho \text{,}  \label{ce10}
\end{equation}
\begin{equation}
nk_BT=\int d{\bf v}^A\frac 12\left( {\bf v}^A-{\bf u}\right) ^2m^Af^A+\int d%
{\bf v}^B\frac 12{({\bf v}}^B-{\bf u)}^2m^Bf^B\text{,}  \label{ce11}
\end{equation}
\begin{equation}
e_{\text{int}}\text{(}P_0\text{)}=nk_BT\text{,}  \label{ce12}
\end{equation}
where $n$, $\rho $, ${\bf u}$, $T$, $e_{\text{int}}$, $P_0$ are the
total number density, total mass density, barycentric velocity, 
mean temperature, total internal energy, and total hydrostatic pressure,
respectively. It is easy to find the following relations, 
\begin{equation}
T=\frac 1n(n^AT^A+n^BT^B)+\frac{\rho ^A\rho ^B}{2nk_B\rho }({\bf u}^A-{\bf u}%
^B)^2\text{,}  \label{ce13a}
\end{equation}
\begin{equation}
P_0=P_0^A+P_0^B+\frac{\rho ^A\rho ^B}{2\rho }({\bf u}^A-{\bf u}^B)^2\text{.}
\label{ce13b}
\end{equation}
Here three sets of hydrodynamic quantities [($\rho ^A$, ${\bf u}^A$, $T^A$),
($\rho ^B$, ${\bf u}^B$, $T^B$) and ($\rho $, ${\bf u}$, $T$)] are
involved.
If assume that the two components are in local equilibrium,
implying that  
 $T^A$, $T^B$, $T$ can be replaced by $T^{(0)}$
 and 
${\bf u}^A$, ${\bf u}^B$, ${\bf u}$ can be replaced by ${\bf
u}^{(0)}$  
in the definitions of $f^{A\left( 0\right) }$
, $f^{B\left( 0\right) }$
 and $f^{AB\left( 0\right) }$,
we arrive at the one-fluid theory and Eq. (\ref{ce13b})
recovers Dalton's law\cite{Dalton}, where $T^{(0)}$ and ${\bf u}^{(0)}$
are the temperature and the velocity of the system in the complete
equilibrium. It is clear that the one-fluid theory is 
conditionally valid. 
If the differences among $T^A$, $T^B$, $T$ and/or 
among ${\bf u}^A$, ${\bf u}^B$, ${\bf u}$ are not small, the above 
replacements result in large errors. 
Since each set of the hydrodynamic quantities can be
described by the other two sets,  
in such cases, a two-fluid theory is preferable. 
Without loss of generality, we require the
description to be dependent on ($\rho ^A$, ${\bf u}^A$, $T^A$) 
and ($\rho ^B$, ${\bf u}^B$, $T^B$)\cite{foot1}.

A key point to complete the two-fluid kinetic description is how to
calculate the local Maxwellian $f^{AB(0)}$ ($%
f^{BA(0)} $). Within Sirovich's original treatments, it is Taylor expanded
around $f^{A(0)}$ ($f^{B(0)}$) to the first order of flow velocity and
temperature\cite{PF5908}. This treatment is reasonable when the hydrodynamic
properties of the two components are nearly symmetric, i.e., $\left| {\bf u}-%
{\bf u}^A\right| \approx \left| {\bf u}-{\bf u}^B\right| $, $\left|
T-T^A\right| \approx \left| T-T^B\right| $. To make a general
theory working also for asymmetric systems where the hydrodynamic 
properties of the two components are greatly different, we introduce 
the reference distribution function in a general way and do the 
Taylor expansion around it.

For $f^{AB(0)}$, we choose the reference distribution function 
 $g^{A\left(0\right) }$ as 
\begin{equation}
g^{A\left( 0\right) }=\frac{n^A}{2\pi \Theta ^{Ar}}\exp \left[ -\frac{\left( 
{\bf v}^A-{\bf u}^{Ar}\right) ^2}{2\Theta ^{Ar}}\right] \text{,}  \label{rf1}
\end{equation}
where the second superscript ``r'' means ``reference'' and the corresponding
quantities are the reference hydrodynamic quantities which take values in
the following way, 
\begin{equation}
\Theta ^{Ar}=\frac{k_BT^{Ar}}{m^A}\text{,}  \label{rf2}
\end{equation}
\begin{equation}
{\bf u}^{Ar}=\left\{ 
\begin{array}{ll}
{\bf u}^A & \text{( if }\left| {\bf u}-{\bf u}^A\right| \leq \left| {\bf u}-%
{\bf u}^B\right| \text{ )} \\ 
{\bf u}^B & \text{( if }\left| {\bf u}-{\bf u}^A\right| >\left| {\bf u}-{\bf %
u}^B\right| \text{ )}
\end{array}
\right. \text{, }  \label{rf3}
\end{equation}
\begin{equation}
T^{Ar}=\left\{ 
\begin{array}{ll}
T^A & \text{( if }\left| T-T^A\right| \leq \left| T-T^B\right| \text{ )} \\ 
T^B & \text{( if }\left| T-T^A\right| >\left| T-T^B\right| \text{ )}
\end{array}
\right. \text{ .}  \label{rf4}
\end{equation}
Let us make the solutions more explicit. Firstly for ${\bf u}$, from Eq.(%
\ref{ce10}) we have 
\begin{equation}
\left| \frac{{\bf u}-{\bf u}^A}{{\bf u}-{\bf u}^B}\right| =\frac{\rho ^B}{%
\rho ^A}=\frac{n^Bm^B}{n^Am^A}\text{.}  \label{rf5}
\end{equation}
Then for $T$, from Eq. (\ref{ce13a}), when $T^A>T^B$ we have 
\begin{eqnarray}
\left| \frac{T-T^A}{T-T^B}\right| &=&\left| \frac{\frac{\rho ^B\rho ^A}{%
2nk_B\rho }({\bf u}^B-{\bf u}^A)^2-\frac{n^B}n(T^A-T^B)}{\frac{\rho ^B\rho ^A%
}{2nk_B\rho }({\bf u}^B-{\bf u}^A)^2+\frac{n^A}n(T^A-T^B)}\right|  \nonumber
\\
&<&1\text{;}  \label{rf6}
\end{eqnarray}
when $T^A<T^B$ we have 
\begin{eqnarray}
\left| \frac{T-T^A}{T-T^B}\right| &=&\left| \frac{\frac{\rho ^B\rho ^A}{%
2nk_B\rho }({\bf u}^B-{\bf u}^A)^2+\frac{n^B}n(T^B-T^A)}{\frac{\rho ^B\rho ^A%
}{2nk_B\rho }({\bf u}^B-{\bf u}^A)^2-\frac{n^A}n(T^B-T^A)}\right|  \nonumber
\\
&>&1.  \label{rf7}
\end{eqnarray}
Considering together (\ref{rf3})-(\ref{rf7}) gives 
\begin{equation}
{\bf u}^{Ar}=\left\{ 
\begin{array}{ll}
{\bf u}^A & \text{( if }\rho ^A\geq \rho ^B\text{ )} \\ 
{\bf u}^B & \text{( if }\rho ^A<\rho ^B\text{ )}
\end{array}
\right. \text{ ,}  \label{rf8}
\end{equation}
\begin{equation}
T^{Ar}=\left\{ 
\begin{array}{ll}
T^A & \text{( if }T^A\geq T^B\text{ )} \\ 
T^B & \text{ (if }T^A<T^B\text{ )}
\end{array}
\right. \text{ .}  \label{rf9}
\end{equation}
In the case of ${\bf u}^{Ar}={\bf u}^A$ and $T^{Ar}=T^A$,  
$g^{A\left( 0\right) }$ gets back to $f^{A\left( 0\right) }$.

Both of $\rho ^A$ and $T^A$ are local quantities. Their values are functions
of position and time. It is possible for such a phenomenon, $\rho ^A(%
{\bf r}_1,t)>\rho ^B({\bf r}_1,t)$ but $\rho ^A({\bf r}_2,t)<\rho ^B({\bf r}%
_2,t)$, to occur, where ${\bf r}_1$ and ${\bf r}_2$ are two different
positions in the system. While in a theory it is not convenient to use the
reference state in such a way: ${\bf u}^{Ar}({\bf r}_1,t)={\bf u}^A({\bf r}%
_1,t)$ and ${\bf u}^{Ar}({\bf r}_2,t)={\bf u}^B({\bf r}_2,t)$. Instead, we
prefer to use one of the two possibilities, ${\bf u}^{Ar}({\bf r},t)={\bf u}%
^A({\bf r},t)$ or ${\bf u}^{Ar}({\bf r},t)={\bf u}^B({\bf r},t)$, in the
whole system, where ${\bf r}$ is an arbitrary position in the system. For $%
T^{Ar}$ we have the same preference. To that aim, $\rho ^A$,$\rho ^B$,$T^A$
and $T^B$ in the criteria (\ref{rf8}) and (\ref{rf9}) are replaced by their
spacially averaged values, $\bar{\rho}^A$, $\bar{\rho}^B$, $\bar{T}^A$ and $\bar{T}^B$,
respectively. This treatment is reasonable from a statistical sense.

\subsection{Kinetic models for symmetric systems}

For systems with
 $\bar{\rho}^A\approx \bar{\rho}^B$ and $\bar{T}^A\approx \bar{%
T}^B$, we can use $g^{A\left( 0\right) }=f^{A\left( 0\right) }$, $g^{B\left(
0\right) }=f^{B\left( 0\right) }$, i.e, Sirovich's kinetic theory. In this
case the equations for the two components are symmetric. 
The cross-collision term in (\ref{ce1}) becomes 
\begin{eqnarray}
\text{ }J^{AB} &=&-\frac 1{\tau ^{AB}}\left[ f^A-f^{A(0)}\right]  \nonumber
\\
&&-\frac{f^{A(0)}}{\Theta ^A}\left\{ \mu _D^A\left( {\bf v}^A-{\bf u}%
^A\right) \cdot ({\bf u}^A-{\bf u}^B)+\mu _T^A\left[ \frac{\left( {\bf v}^A-%
{\bf u}^A\right) ^2}{2\Theta ^A}-1\right] (T^A-T^B)\right.  \nonumber \\
&&\left. -M^A\left[ \frac{\left( {\bf v}^A-{\bf u}^A\right) ^2}{2\Theta ^A}%
-1\right] ({\bf u}^A-{\bf u}^B)^2\right\}  \label{ce16}
\end{eqnarray}
where 
\begin{equation}
\mu _D^A=\frac{\rho ^B}{\tau ^{AB}\rho }\text{, }\mu _T^A=\frac{k_Bn^B}{\tau
^{AB}nm^A}\text{, }M^A=\frac{n^A\rho ^B}{2\tau ^{AB}n\rho }\text{.}
\label{ce17}
\end{equation}
If we concern the hydrodynamics only up to the NSE level, $%
f^A $ ($f^B$) in the force term can be replaced by $f^{A\left( 0\right) }$ ($%
f^{B(0)}$). The BGK model (\ref{ce1}-\ref{ce4a}) can be rewritten as 
\begin{equation}
\partial _tf^A+{\bf v}^A{\bf \cdot }\frac \partial {\partial {\bf r}}f^A-%
{\bf a}^A\cdot \frac{\left( {\bf v}^A-{\bf u}^A\right) }{\Theta ^A}%
f^{A\left( 0\right) }=Q^{AA}+Q^{AB}\text{,}  \label{ce21}
\end{equation}
\begin{equation}
\partial _tf^B+{\bf v}^B{\bf \cdot }\frac \partial {\partial {\bf r}}f^B-%
{\bf a}^B\cdot \frac{\left( {\bf v}^B-{\bf u}^B\right) }{\Theta ^B}%
f^{B\left( 0\right) }=Q^{BB}+Q^{BA}\text{,}  \label{ce21b}
\end{equation}
where 
\begin{equation}
Q^{AA}=-\frac 1{\tau ^A}\left[ f^A-f^{A(0)}\right] \text{, }\frac 1{\tau ^A}=%
\frac 1{\tau ^{AA}}+\frac 1{\tau ^{AB}}  \label{ce22}
\end{equation}
\begin{eqnarray}
\text{ }Q^{AB} &=&-\frac{f^{A(0)}}{\Theta ^A}\left\{ \mu _D^A\left( {\bf v}%
^A-{\bf u}^A\right) \cdot ({\bf u}^A-{\bf u}^B)+\mu _T^A\left[ \frac{\left( 
{\bf v}^A-{\bf u}^A\right) ^2}{2\Theta ^A}-1\right] (T^A-T^B)\right. 
\nonumber \\
&&\left. -M^A\left[ \frac{\left( {\bf v}^A-{\bf u}^A\right) ^2}{2\Theta ^A}%
-1\right] ({\bf u}^A-{\bf u}^B)^2\right\} \text{;}  \label{ce23}
\end{eqnarray}
the expressions of $Q^{BB}$ and $Q^{BA}$ are obtained from
Eqs. (\ref{ce22}) and (\ref{ce23}) via formal replacements of the
superscripts $A$ and $B$.
In the isothermal case, 
$T^A=T^B=T$,  the expression of $Q^{AB}$ is simplified as 
\begin{equation}
\text{ }Q^{AB}=-\frac{f^{A(0)}}{\Theta ^A}\mu _D^A\left( {\bf v}^A-{\bf u}%
^A\right) \cdot ({\bf u}^A-{\bf u}^B)\text{.}  \label{ce24}
\end{equation}
For the convenience of description, the kinetic model with (\ref{ce21})-(\ref
{ce23}) is referred to as kinetic model A; the one with (\ref{ce21})-(\ref
{ce22}),(\ref{ce24}) is referred to as kinetic model B.

\subsection{Kinetic models for asymmetric systems}

\subsubsection{Kinetic model C: for isothermal systems with $\bar{\rho}^A>%
\bar{\rho}^B$}

For such a system, $g^{A\left( 0\right) }=f^{A\left(
0\right) }$, and 
\begin{equation}
g^{B\left( 0\right) }=\frac{\rho ^B}{2\pi k_BT}\exp \left[ -\frac{m^B\left( 
{\bf v}^B-{\bf u}^A\right) ^2}{2k_BT}\right] \text{,}  \label{kc1}
\end{equation}
\begin{eqnarray}
f^{BA\left( 0\right) } &=&g^{B\left( 0\right) }+g^{B\left( 0\right) }\left( 
\frac{m^B}{k_BT}\right) \left( {\bf v}^B-{\bf u}^A\right) \cdot \left( {\bf u%
}-{\bf u}^A\right)  \nonumber \\
&=&g^{B\left( 0\right) }\left[ 1+\frac{\rho ^B}\rho \frac{\left( {\bf v}^B-%
{\bf u}^A\right) }{\Theta ^B}\cdot ({\bf u}^B-{\bf u}^A)\right] \text{.}
\label{kc2}
\end{eqnarray}
Thus, within kinetic model C, 
\begin{equation}
Q^{AA}=-\frac 1{\tau ^A}\left[ f^A-f^{A\left( 0\right) }\right] 
\text{,}  \label{kc3}
\end{equation}
\begin{equation}
Q^{BB}=-\frac 1{\tau ^{BB}}\left[ f^B-f^{B\left( 0\right) }\right] -\frac 1{%
\tau ^{BA}}\left[ f^B-g^{B\left( 0\right) }\right] \text{,}  \label{kc4}
\end{equation}
\begin{equation}
Q^{AB}=-\frac{f^{A\left( 0\right) }}{\Theta ^A}\mu _D^A\left( {\bf v}^A-{\bf %
u}^A\right) \cdot ({\bf u}^A-{\bf u}^B)\text{,}  \label{kc5}
\end{equation}
\begin{equation}
Q^{BA}=-\frac{g^{B\left( 0\right) }}{\Theta ^B}\mu _D^{B*}\left( {\bf v}^B-%
{\bf u}^A\right) \cdot ({\bf u}^A-{\bf u}^B)\text{,}  \label{kc6}
\end{equation}
where 
\begin{equation}
\mu _D^{B*}=\frac{\rho ^B}{\tau ^{BA}\rho }\text{.}  \label{muDB2}
\end{equation}

\subsubsection{Kinetic model D: for systems with $\bar{\rho}^A>\bar{\rho}^B$
and $\bar{T}^A>\bar{T}^B$}

The references are  $g^{A\left(0\right) }=f^{A\left( 0\right) }$ and 
\begin{equation}
g^{B\left( 0\right) }=\frac{\rho ^B}{2\pi k_BT^A}\exp \left[ -\frac{%
m^B\left( {\bf v}^B-{\bf u}^A\right) ^2}{2k_BT^A}\right] \text{.}
\label{kd01}
\end{equation}
Since 
\begin{eqnarray}
f^{BA\left( 0\right) } &=&g^{B\left( 0\right) }\left\{ 1+\frac{\rho ^B}\rho 
\frac{\left( {\bf v}^B-{\bf u}^A\right) }{\Theta ^{Br}}\cdot ({\bf u}^B-{\bf %
u}^A)\right.  \nonumber \\
&&\left. +\left[ \frac{\left( {\bf v}^B-{\bf u}^A\right) ^2}{2\Theta ^{Br}T^A%
}-\frac 1{T^A}\right] \left[ \frac{n^B}n\left( T^B-T^A\right) +\frac{\rho
^B\rho ^A}{2nk_B\rho }({\bf u}^B-{\bf u}^A)^2\right] \right\} \text{,}
\label{kd02}
\end{eqnarray}
within kinetic model D  
\begin{equation}
Q^{AA}=-\frac 1{\tau ^A}\left[ f^A-f^{A\left( 0\right) }\right] 
\text{,}  \label{kd1}
\end{equation}
\begin{equation}
Q^{BB}=-\frac 1{\tau ^{BB}}\left[ f^B-f^{B\left( 0\right) }\right] -\frac 1{%
\tau ^{BA}}\left[ f^B-g^{B\left( 0\right) }\right] \text{,}  \label{kd2}
\end{equation}
\begin{eqnarray}
Q^{AB} &=&-\frac{f^{A\left( 0\right) }}{\Theta ^A}\left\{ \mu _D^A\left( 
{\bf v}^A-{\bf u}^A\right) \cdot ({\bf u}^A-{\bf u}^B)+\mu _T^A\left[ \frac{%
\left( {\bf v}^A-{\bf u}^A\right) ^2}{2\Theta ^A}-1\right] (T^A-T^B)\right. 
\nonumber \\
&&\left. -M^A\left[ \frac{\left( {\bf v}^A-{\bf u}^A\right) ^2}{2\Theta ^A}%
-1\right] ({\bf u}^A-{\bf u}^B)^2\right\} \text{,}  \label{kd3}
\end{eqnarray}
\begin{eqnarray}
Q^{BA} &=&-\frac{g^{B\left( 0\right) }}{\Theta ^{Br}}\left\{ \mu
_D^{B*}\left( {\bf v}^B-{\bf u}^A\right) \cdot ({\bf u}^A-{\bf u}^B)+\mu
_T^{B*}\left[ \frac{\left( {\bf v}^B-{\bf u}^A\right) ^2}{2\Theta ^{Br}}%
-1\right] \left( T^A-T^B\right) \right.  \nonumber \\
&&\left. -M^B\left[ \frac{\left( {\bf v}^B-{\bf u}^A\right) ^2}{2\Theta ^{Br}%
}-1\right] ({\bf u}^B-{\bf u}^A)^2\right\} \text{,}  \label{kd4}
\end{eqnarray}
where 
\begin{equation}
\mu _T^{B*}=k_B\frac{n^B}{\tau ^{BA}m^Bn}\text{.}  \label{mc7}
\end{equation}

\subsubsection{Kinetic model E: for systems with $\bar{\rho}^A>\bar{\rho}^B$
and $\bar{T}^A<\bar{T}^B$}

In this case, the reference velocity and reference temperature for both $%
f^{AB\left( 0\right) }$ and $f^{BA\left( 0\right) }$ are ${\bf u}^A$ and $%
T^B $, respectively. 
\begin{equation}
g^{A\left( 0\right) }=\frac{\rho ^A}{2\pi k_BT^B}\exp \left[ -\frac{%
m^A\left( {\bf v}^A-{\bf u}^A\right) ^2}{2k_BT^B}\right] \text{,}
\label{ke1}
\end{equation}
\begin{equation}
g^{B\left( 0\right) }=\frac{\rho ^B}{2\pi k_BT^B}\exp \left[ -\frac{%
m^B\left( {\bf v}^B-{\bf u}^A\right) ^2}{2k_BT^B}\right] \text{.}
\label{ke2}
\end{equation}
Within the kinetic model E 
\begin{equation}
\text{ }Q^{AA}=-\frac 1{\tau ^{AA}}\left[ f^A-f^{A\left( 0\right) }\right] -%
\frac 1{\tau ^{AB}}\left[ f^A-g^{A\left( 0\right) }\right] \text{,}
\label{ke3}
\end{equation}
\begin{equation}
Q^{BB}=-\frac 1{\tau ^{BB}}\left[ f^B-f^{B\left( 0\right) }\right] -\frac 1{%
\tau ^{BA}}\left[ f^B-g^{B\left( 0\right) }\right] \text{,}  \label{ke4}
\end{equation}
\begin{eqnarray}
\text{ }Q^{AB} &=&-\frac{g^{A\left( 0\right) }}{\Theta ^{Ar}}\left\{ \mu
_D^A\left( {\bf v}^A-{\bf u}^A\right) \cdot ({\bf u}^A-{\bf u}^B)+
\frac{k_B n^A}{\tau^{AB}n m^A}
\left[ \frac{\left( {\bf v}^A-{\bf u}^A\right) ^2}{2\Theta ^{Ar}}%
-1\right] (T^B-T^A)\right.  \nonumber \\
&&\left. -M^A\left[ \frac{\left( {\bf v}^A-{\bf u}^A\right) ^2}{2\Theta ^{Ar}%
}-1\right] ({\bf u}^A-{\bf u}^B)^2\right\} \text{,} \label{ke5}
\end{eqnarray}
\begin{eqnarray}
Q^{BA} &=&-\frac{g^{B\left( 0\right) }}{\Theta ^B}\left\{ \mu _D^{B*}\left( 
{\bf v}^B-{\bf u}^A\right) \cdot ({\bf u}^A-{\bf u}^B)+
\frac{k_B n^A}{\tau^{BA}n m^B}
\left[ 
\frac{\left( {\bf v}^B-{\bf u}^A\right) ^2}{2\Theta ^B}-1\right] \left(
T^B - T^A\right) \right.  \nonumber \\
&&\left. -M^B\left[ \frac{\left( {\bf v}^B-{\bf u}^A\right) ^2}{2\Theta ^B}%
-1\right] ({\bf u}^B-{\bf u}^A)^2\right\} \text{.} \label{ke6}
\end{eqnarray}


\subsection{Hydrodynamics and diffusion}

\subsubsection{Hydrodynamics}

A connection between a kinetic model and corresponding hydrodynamics
is the Chapman-Enskog analysis\cite{Chap}. All above kinetic models
contribute to (i) the same continuity equation at the Euler and the NSE
levels, 
\begin{eqnarray}
\frac{\partial \rho ^A}{\partial t}+\frac \partial {\partial r_\alpha }%
\left( \rho ^Au_\alpha ^A\right) &=&0\text{,}  \label{continuity1} \\
\frac{\partial \rho ^B}{\partial t}+\frac \partial {\partial r_\alpha }%
\left( \rho ^Bu_\alpha ^B\right) &=&0 \text{;}  \label{continuity2}
\end{eqnarray}
(ii) the same Euler momentum equations, 
\begin{equation}
\frac \partial {\partial t}\left( \rho ^Au_\alpha ^A\right) {\bf +}\frac %
\partial {\partial r_\beta }\left( \rho ^Au_\alpha ^Au_\beta ^A\right) +%
\frac{\partial P_0^A}{\partial r_\alpha }-\rho ^Aa_\alpha ^A+\frac{j_\alpha
^{AB}}{\tau ^{AB}}=0\text{,}  \label{Ema}
\end{equation}
\begin{equation}
\frac \partial {\partial t}\left( \rho ^Bu_\alpha ^B\right) {\bf +}\frac %
\partial {\partial r_\beta }\left( \rho ^Bu_\alpha ^Bu_\beta ^B\right) +%
\frac{\partial P_0^B}{\partial r_\alpha }-\rho ^Ba_\alpha ^B+\frac{j_\alpha
^{BA}}{\tau ^{BA}}=0\text{;}  \label{Emb}
\end{equation}
and (iii) the same NSE momentum equation for component $A$, 
\begin{equation}
\frac \partial {\partial t}\left( \rho ^Au_\alpha ^A\right) {\bf +}\frac %
\partial {\partial r_\beta }\left( \rho ^Au_\alpha ^Au_\beta ^A\right) +%
\frac{\partial P_{\alpha \beta }^A}{\partial r_\beta }-\rho ^Aa_\alpha ^A+%
\frac{j_\alpha ^{AB}}{\tau ^{AB}}=0\text{,}  \label{Nma}
\end{equation}
where 
\begin{equation}
j_\alpha ^{AB}=-j_\alpha ^{BA}=\frac{\rho ^A\rho ^B}\rho \left( u_\alpha
^A-u_\alpha ^B\right)  \label{dflx}
\end{equation}
describes the momentum transferred from component $A$ to $B$, and it is
also the diffusion flux density which will be clear from a later 
equation (\ref{da1}); 
\begin{equation}
P_{\alpha \beta }^A=P_0^A\delta _{\alpha \beta }-\Pi _{\alpha \beta }^A
\label{PrssA}
\end{equation}
is the stress tensor, 
\begin{equation}
\Pi _{\alpha \beta }^A=\eta ^A\left( \frac{\partial u_\alpha ^A}{\partial
r_\beta }+\frac{\partial u_\beta ^A}{\partial r_\alpha }-\frac{\partial
u_\gamma ^A}{\partial r_\gamma }\delta _{\alpha \beta }\right)
\label{stressA}
\end{equation}
is the viscous stress tensor, and 
\begin{equation}
\eta ^A=P^A_0 \tau ^A  \label{etaa}
\end{equation}
is the viscosity.

Moldes A and B contributes to symmetric hydrodynamics for the two
components. The Euler energy equation of model A for component $A$ reads,
\begin{equation}
\frac{\partial e^A}{\partial t}{\bf +}\frac \partial {\partial r_\alpha }%
\left[ \left( e^A+P_0^A\right) u_\alpha ^A\right] -\rho ^Aa_\alpha
^Au_\alpha ^A{\bf +}\frac 1{\tau ^{AB}}\left( u_\alpha ^Aj_\alpha
^{AB}+q^{AB}-\frac{n^A\rho }{2n\rho ^A\rho ^B}j_\alpha ^{AB}j_\alpha
^{AB}\right) =0\text{,}  \label{Eena}
\end{equation}
where 
\begin{equation}
e^A=e_{\text{int}}^A+\frac 12\rho ^A\left( u^A\right) ^2  \label{etotal}
\end{equation}
is the local total energy, and 
\begin{equation}
q^{AB}=\frac{n^An^B}nk_B\left( T^A-T^B\right)  \label{heat}
\end{equation}
is the heat transfered from component $A$ to $B$.

The NSE momentum equation for component $B$ from model C reads
\begin{equation}
\frac \partial {\partial t}\left( \rho ^Bu_\alpha ^B\right) +\frac \partial {%
\partial r_\beta }\left( \rho ^Bu_\alpha ^Bu_\beta ^B\right) {\bf +}\frac{%
\partial \left( P_{\alpha \beta }^B+\tilde{\Pi}_{\alpha \beta }^B\right) }{%
\partial r_\beta }-\rho ^Ba_\alpha ^B+\frac{j_\alpha ^{BA}}{\tau ^{BA}}=0%
\text{,}  \label{Cmb}
\end{equation}
where the definition of $P_{\alpha \beta }^B$ is similar to that of $%
P_{\alpha \beta }^A$, and 
\begin{equation}
\tilde{\Pi}_{\alpha \beta }^B=-\frac{\tau ^B\rho ^B\left( \rho ^B-\rho
^A\right) }{\tau ^{BA}\rho }\left[ \left( u_\alpha ^B-u_\alpha ^A\right)
\left( u_\beta ^B-u_\beta ^A\right) \right]  \label{astressB}
\end{equation}
is an additional stress tensor due to the asymmetry of densities of the 
two components.

The Euler energy equation of model D for component $A$ is the same as Eq. (%
\ref{Eena}) and for component $B$ reads 
\begin{eqnarray}
&& \frac{\partial e^B}{\partial t}{\bf +}\frac \partial {\partial r_\alpha }%
\left[ \left( e^B+P_0^B\right) u_\alpha ^B\right] -\rho ^Ba_\alpha
^Bu_\alpha ^B  \nonumber \\
&& +\frac 1{\tau ^{BA}}\left\{ \frac{\rho ^B}2\left( u_\alpha ^Bu_\alpha
^B-u_\alpha ^Au_\alpha ^A\right) +\frac{\left( \rho ^B\right) ^2}\rho \left(
u_\alpha ^Au_\alpha ^A-u_\alpha ^Au_\alpha ^B\right) +q^{BA}-\frac{n^A\rho }{%
2n\rho ^A\rho ^B}j_\alpha ^{BA}j_\alpha ^{BA}\right\}  \nonumber \\
&& =0\text{,}  \label{Deb}
\end{eqnarray}
where the definition of $e^B$ is similar to that of $e^A$ and 
\[
q^{BA}=-q^{AB}\text{.} 
\]
The Euler energy equations from kinetic model E are as follows,
\begin{eqnarray}
&& \frac{\partial e^A}{\partial t}+\frac \partial {\partial r_\alpha }%
\left[ \left( e^A+P_0^A\right) u_\alpha ^A\right] -\rho ^Aa_\alpha
^Au_\alpha ^A \nonumber \\
&& +\frac 1{\tau ^{AB}}\left( u_\alpha ^Aj_\alpha^{AB}
+
q^{AB}-
\frac{n^A\rho }{2n\rho ^A\rho ^B}j_\alpha ^{AB}j_\alpha
^{AB}\right)=0 \text{,}  \label{Eea}
\end{eqnarray}
\begin{eqnarray}
&&\frac{\partial e^B}{\partial t}+\frac \partial {\partial r_\alpha }%
\left[ \left( e^B+P_0^B\right) u_\alpha ^B\right] -\rho ^Ba_\alpha
^Bu_\alpha ^B  \nonumber \\
&&+\frac 1{\tau ^{BA}}\left[ \frac{\rho ^B}2\left( u_\alpha ^Bu_\alpha
^B-u_\alpha ^Au_\alpha ^A\right) +\frac{\left( \rho ^B\right) ^2}\rho \left(
u_\alpha ^Au_\alpha ^A-u_\alpha ^Au_\alpha ^B\right) +
q^{BA}-%
\frac{n^A\rho }{2n\rho ^A\rho ^B}j_\alpha ^{BA}j_\alpha ^{BA}\right] 
\nonumber \\
&&=0\text{.}  \label{Eeb}
\end{eqnarray}

\subsubsection{Diffusion}

From the continuity equations (\ref{continuity1})-(\ref{continuity2}) we
have 
\begin{equation}
\frac{\partial \rho }{\partial t}+\frac \partial {\partial r_\alpha }\left(
\rho u_\alpha \right) =0  \label{mt}
\end{equation}
and 
\begin{equation}
\frac{\partial \rho ^A}{\partial t}+\frac \partial {\partial r_\alpha }%
\left( \rho ^Au_\alpha \right) =-\frac{\partial j_\alpha ^{AB}}{\partial
r_\alpha }\text{.}  \label{da1}
\end{equation}
where $j_\alpha ^{AB}$ is given in Eq. (\ref{dflx}) and it is the amount of
the component $A$ transported relative to the component $B$ by diffusion
through unit area in unit time. For the incompressible fluids where $\rho $
is a constant, the continuity equation (\ref{da1}) is equivalent to the
following diffusion-convection equation, 
\begin{equation}
\frac{\partial \varphi }{\partial t}+\frac \partial {\partial r_\alpha }%
\left( \varphi u_\alpha \right) =-\frac \partial {\partial r_\alpha }\left[
\varphi \left( 1-\varphi \right) \left( u_\alpha ^A-u_\alpha ^B\right)
\right] \text{,}  \label{cd3}
\end{equation}
where $\varphi =\rho ^A/\rho $. The diffusion velocity $\left( u_\alpha
^A-u_\alpha ^B\right) $ is determined by the momentum equation. We can find
a simple relation for it in the following case: We consider a binary system
without external forces and where the flow velocities 
$u_\alpha ^A$, $u_\alpha ^B$ are
small and their derivatives can be regarded as higher-order small
quantities. 
 From the momentum Eq. (\ref{Ema}) or (\ref{Nma}),
by neglecting the second and higher-order terms in $u_\alpha ^A$
and/or $u_\alpha ^B$, then using the definition (\ref{ce8a}), we obtain  
\begin{eqnarray}
u_\alpha ^A-u_\alpha ^B &=&-\frac{\rho \tau ^{AB}}{\rho ^A\rho ^B}\frac{%
\partial P_0^A}{\partial r_\alpha }  \label{dv1} \\
&=&-\frac{k_B\tau ^{AB}\rho T^A}{\rho ^A\rho ^B}\frac{\partial n^A}{\partial
r_\alpha }-\frac{k_B\tau ^{AB}\rho }{m^A\rho ^B}\frac{\partial T^A}{\partial
r_\alpha }\text{.}  \label{dv2}
\end{eqnarray}
If further assume the system to be isothermal, the density flux of
component $A$ reads
\begin{equation}
j_\alpha ^A=\rho ^A\left( u_\alpha ^A-u_\alpha \right) =-{\cal D}^A\frac{%
\partial \rho ^A}{\partial r_\alpha }\text{,}  \label{Fick1}
\end{equation}
where 
\begin{equation}
{\cal D}^A=\frac{k_BT\tau ^{AB}}{m^A}  \label{Difa}
\end{equation}
is the diffusivity of component $A$. Eq. (\ref{Fick1}) is Fick's
first law\cite{Web}. From Eqs. (\ref{cd3}) and (\ref{dv2}) we have 
\begin{equation}
\frac{\partial \varphi }{\partial t}+\frac \partial {\partial r_\alpha }%
\left( \varphi u_\alpha \right) = \frac \partial {\partial r_\alpha }\left[ 
{\cal D}^A\frac{\partial \varphi }{\partial r_\alpha }\right] \text{.}
\label{dv3}
\end{equation}
In the case where the barycentric velocity field is zero and $\tau ^{AB}$ is
a constant, the diffusion-convection equation (\ref{dv3}) reduces to Fick's
second law\cite{Web}, 
\begin{equation}
\frac{\partial \varphi }{\partial t}= {\cal D}^A\frac \partial {\partial
r_\alpha }\frac{\partial \varphi }{\partial r_\alpha }\text{.}  \label{dv4}
\end{equation}

Under the present treatment cross-collisions contribute to the viscous
behavior and are responsible for the inter-diffusion as well  as
momentum and heat exchanges between the two components.   
The momentum and heat exchanges between the two
components occur not only at the
Navier-Stokes level but also at the Euler level\cite{foot2}, 
which is different from
the case in the 
one-fluid theory\cite{VictorIJMPC,Dalton}, but consistent with the
two-fluid relations (\ref{ce13a}) and (\ref{ce13b}).

\section{Discrete kinetic models}

\subsection{General description}

Based on the following discrete velocity model, 
\begin{equation}
v_0=0\text{, }{\bf v}_{ki}=v_k\left[ \cos \left( \frac{i\pi }4\right) \text{%
, }\sin \left( \frac{i\pi }4\right) \right] \text{, }i=1\text{, }2\text{, }%
\cdots \text{,}8 \text{,}  \label{i2}
\end{equation}
the kinetic equations read 
\begin{equation}
\frac{\partial f_{ki}^A}{\partial t}{\bf +}v_{ki\alpha }^A\frac{\partial
f_{ki}^A}{\partial r_\alpha }-{\bf a}^A\cdot \frac{\left( {\bf v}_{ki}^A-%
{\bf u}^A\right) }{\Theta ^A}f_{ki}^{A\left( 0\right)
}=Q_{ki}^{AA}+Q_{ki}^{AB}\text{,}  \label{DB1}
\end{equation}
\begin{equation}
\frac{\partial f_{ki}^B}{\partial t}{\bf +}v_{ki\alpha }^B\frac{\partial
f_{ki}^B}{\partial r_\alpha }-{\bf a}^B\cdot \frac{\left( {\bf v}_{ki}^B-%
{\bf u}^B\right) }{\Theta ^B}f_{ki}^{B\left( 0\right)
}=Q_{ki}^{BB}+Q_{ki}^{BA}\text{,}  \label{DB1b}
\end{equation}
where subscript $k$ indicates the $k$-th group of particle velocities and $i$
indicates the direction of the particle speed. The DVM (\ref{i2})
is isotropic up to its seventh rank tensors\cite{PRE6736306}.
The discrete kinetic model, (\ref{DB1})-(\ref{DB1b}), 
is required to recover the same hydrodynamic equations as those of its 
continuous version. This requirement is used to formulate the 
multispeed-discrete-velocity kinetic models.

\subsection{Models for isothermal and compressible Navier-Stokes
equations}

\subsubsection{Discrete-velocity kinetic model B}

Due to the symmetry of the two components, we show results only for the
component $A$, 
\begin{equation}
Q_{ki}^{AA}=-\frac 1{\tau ^A}\left[ f_{ki}^A-f_{ki}^{A(0)}\right] \text{,}
\label{DB3}
\end{equation}
\begin{equation}
\text{ }Q_{ki}^{AB}=-\frac{f_{ki}^{A(0)}}{\Theta ^A}\mu _D^A\left( {\bf v}%
_{ki}^A-{\bf u}^A\right) \cdot ({\bf u}^A-{\bf u}^B)\text{,}  \label{DB4}
\end{equation}
\begin{equation}
f_{ki}^{A(0)}=n^A\left( \frac{m^A}{2\pi k_BT^A}\right) \exp \left[ -\frac{%
m^A\left( {\bf v}_{ki}^A-{\bf u}^A\right) ^2}{2k_BT^A}\right] \text{.}
\label{DB5}
\end{equation}
The Chapman-Enskog analysis shows that, to get the isothermal NSE 
equations, (\ref{continuity1}) and (\ref{Nma}), the following requirements
on the discrete equilibrium distribution function, 
\begin{equation}
\sum_{ki}f_{ki}^{A(0)}=n^A\text{,}  \label{rti1}
\end{equation}
\begin{equation}
\sum_{ki}{\bf v}_{ki}^Af_{ki}^{A(0)}=n^A{\bf u}^A\text{,}  \label{rti2}
\end{equation}
\begin{equation}
\sum_{ki}m^Av_{ki\alpha }^Av_{ki\beta }^Af_{ki}^{A(0)}=P^A_0 \delta _{\alpha
\beta }+\rho ^Au_\alpha ^Au_\beta ^A\text{,}  \label{rti7}
\end{equation}
\begin{equation}
\sum_{ki}m^Av_{ki\alpha }^Av_{ki\beta }^Av_{ki\gamma
}^Af_{ki}^{A(0)}=P^A_0 \left( u_\gamma ^A\delta _{\alpha \beta }+u_\alpha
^A\delta _{\beta \gamma }+u_\beta ^A\delta _{\gamma \alpha }\right) +\rho
^Au_\alpha ^Au_\beta ^Au_\gamma ^A\text{,}  \label{rti9}
\end{equation}
are necessary and also sufficient.

The requirement (\ref{rti9}) contains the third order of the flow
velocity 
$u^A$. So it is reasonable to expand $f_{ki}^{A(0)}$ in the polynomial 
form to the third order in the flow velocity,
\begin{eqnarray}
f_{ki}^{A(0)} &=&n^AF_k^A\left\{ \left[ 1-\frac{(u^A)^2}{2\Theta ^A}\right] +%
\frac 1{\Theta ^A}\left( 1-\frac{(u^A)^2}{2\Theta ^A}\right) v_{ki\xi
}^Au_\xi ^A+\frac 1{2\left( \Theta ^A\right) ^2}v_{ki\xi }^Av_{ki\pi
}^Au_\xi ^Au_\pi ^A\right.  \nonumber \\
&&\left. +\frac 1{6\left( \Theta ^A\right) ^3}v_{ki\xi }^Av_{ki\pi
}^Av_{ki\eta }^Au_\xi ^Au_\pi ^Au_\eta ^A\right\} +\cdots  \text{,} \label{feq}
\end{eqnarray}
where 
\begin{equation}
F_k^A=\frac{1}{2\pi\Theta^A}\exp[-\frac{(v_k^A)^2}{2\Theta^A}]. \label{Fk}
\end{equation}
The left-hand side of Eq. (\ref{rti9}) with the truncated $f_{ki}^{A(0)}$
has sixth rank tensor in particle velocity ${\bf v}^A$.
Therefore, to recover the correct hydrodynamical equations, the 
based DVM should be isotropic up to its sixth rank tensor. 
DVM (\ref{i2}) satisfies the need.

To satisfy (\ref{rti1}), we require 
\begin{equation}
\sum_{ki}F_k^A=1\text{,}  \label{rti1r1}
\end{equation}
\begin{equation}
\sum_{ki}F_k^Av_{ki\xi }^Av_{ki\pi }^Au_\xi ^Au_\pi ^A=\Theta ^A\left(
u^A\right) ^2\text{,}  \label{rti1r2}
\end{equation}
To satisfy (\ref{rti2}), we require 
\begin{equation}
\sum_{ki}F_k^Av_{ki\alpha }^Av_{ki\xi }^Au_\xi ^A=\Theta ^Au_\alpha ^A
\label{rti2r1}
\end{equation}
\begin{equation}
\sum_{ki}F_k^Av_{ki\alpha }^Av_{ki\xi }^Av_{ki\pi }^Av_{ki\eta }^Au_\xi
^Au_\pi ^Au_\eta ^A=3\left( \Theta ^A\right) ^2\left( u^A\right) ^2u_\alpha
^A\text{.}  \label{rti2r2}
\end{equation}
To satisfy (\ref{rti7}), we require 
\begin{equation}
\sum_{ki}F_k^Av_{ki\alpha }^Av_{ki\beta }^A=\Theta ^A\delta _{\alpha \beta }
\label{rti7r1}
\end{equation}
\begin{equation}
\sum_{ki}F_k^Av_{ki\alpha }^Av_{ki\beta }^Av_{ki\xi }^Av_{ki\pi }^Au_\xi
^Au_\pi ^A=\left( \Theta ^A\right) ^2\left[ \left( u^A\right) ^2\delta
_{\alpha \beta }+2u_\alpha ^Au_\beta ^A\right]  \label{rti7r2}
\end{equation}
To satisfy (\ref{rti9}), we require 
\begin{equation}
\sum_{ki}F_k^Av_{ki\alpha }^Av_{ki\beta }^Av_{ki\gamma }^Av_{ki\xi }^Au_\xi
=\left( \Theta ^A\right) ^2\left( u_\alpha ^A\delta _{\beta \gamma }+u_\beta
^A\delta _{\gamma \alpha }+u_\gamma ^A\delta _{\alpha \beta }\right)
\label{rti9r1}
\end{equation}
\begin{eqnarray}
&&\sum_{ki}F_k^Av_{ki\alpha }^Av_{ki\beta }^Av_{ki\gamma }^Av_{ki\xi
}^Av_{ki\pi }^Av_{ki\eta }^Au_\xi ^Au_\pi ^Au_\eta ^A  \nonumber \\
&=&3\left( \Theta ^A\right) ^3\left( u^A\right) ^2\left( u_\alpha ^A\delta
_{\beta \gamma }+u_\beta ^A\delta _{\gamma \alpha }+u_\gamma ^A\delta
_{\alpha \beta }\right) +6\left( \Theta ^A\right) ^3u_\alpha ^Au_\beta
^Au_\gamma ^A  \label{rti92a}
\end{eqnarray}

If further consider the isotropic properties of the discrete velocity model,
the above 8 requirements reduce to the following four ones.
Requirement (\ref{rti1r1}) gives 
\begin{equation}
\sum_{ki}F_k^A=1\text{.}  \label{rei1}
\end{equation}
Requirements (\ref{rti1r2}), (\ref{rti2r1}), (\ref{rti7r1}) give 
\begin{equation}
\sum_kF_k^A\left( v_k^A\right) ^2=\frac{\Theta ^A}4\text{.}  \label{rei2}
\end{equation}
Requirements (\ref{rti2r2}), (\ref{rti7r2}), (\ref{rti9r1}) give 
\begin{equation}
\sum_kF_k^A\left( v_k^A\right) ^4=\left( \Theta ^A\right) ^2\text{.}
\label{rei3}
\end{equation}
Requirement (\ref{rti92a}) give 
\begin{equation}
\sum_kF_k^A\left( v_k^A\right) ^6=6\left( \Theta ^A\right) ^3\text{.}
\label{rei4}
\end{equation}
To satisfy the above four requirements, four different particle
velocities are
sufficient. We choose a zero speed, $v_0^A=0$, and other three nonzero ones, 
$v_k^A$ ($k=1$, $2$, $3$). From (\ref{rei2})-(\ref{rei4}) it is easy to find
the following solution, 
\begin{equation}
F_k^A=\frac{\Psi _k^A}{\Phi _k^A}  \label{NFk1}
\end{equation}
\begin{equation}
\Psi _k^A=\Theta ^A\left\{ \left( v_{k+1}^Av_{k+2}^A\right) ^2-4\Theta
^A\left[ \left( v_{k+1}^A\right) ^2+\left( v_{k+2}^A\right) ^2\right]
+24\left( \Theta ^A\right) ^2\right\}  \label{NFk2}
\end{equation}
\begin{equation}
\Phi _k^A=4\left( v_k^A\right) ^2\left\{ \left( v_{k+1}^Av_{k+2}^A\right)
^2-\left( v_k^A\right) ^2\left[ \left( v_{k+1}^A\right) ^2+\left(
v_{k+2}^A\right) ^2\right] +\left( v_k^A\right) ^4\right\}  \label{NFk3}
\end{equation}
where $k=1,2,3$ and $v_4^A=v_1^A$, $v_5^A=v_2^A$. From (\ref{rei1}) we get 
\begin{equation}
F_0^A=1-8\sum_{k=1}^3F_k^A\text{.}  \label{F0N}
\end{equation}

\subsubsection{Discrete-velocity kinetic model C}

The description for component $A$ is the same as that in
 discrete-velocity kinetic model B. For component $B$,
\begin{equation}
Q_{ki}^{BB}=-\frac 1{\tau ^{BB}}\left[ f_{ki}^B-f_{ki}^{B\left( 0\right)
}\right] -\frac 1{\tau ^{BA}}\left[ f_{ki}^B-g_{ki}^{B\left( 0\right)
}\right] \text{,}  \label{DC1}
\end{equation}
\begin{equation}
Q_{ki}^{BA}=-\frac{g_{ki}^{B\left( 0\right) }}{\Theta ^B}\mu _D^{B*}\left( 
{\bf v}_{ki}^B-{\bf u}^A\right) \cdot ({\bf u}^A-{\bf u}^B)\text{.}
\label{DC2}
\end{equation}
$f_{ki}^{A\left( 0\right) }$ and $f_{ki}^{B\left( 0\right) }$ are
formulated in the same as those in discrete model B. Additionally, 
$g_{ki}^{B\left( 0\right) }$ should be formulated in a similar way. Due to
similar reasons, $g_{ki}^{B\left( 0\right) }$ is expanded as
\begin{eqnarray}
g_{ki}^{B(0)} &=&n^BG_k^B\left\{ \left[ 1-\frac{(u^A)^2}{2\Theta ^B}\right] +%
\frac 1{\Theta ^B}\left( 1-\frac{(u^A)^2}{2\Theta ^B}\right) v_{ki\xi
}^Bu_\xi ^A+\frac 1{2\left( \Theta ^B\right) ^2}v_{ki\xi }^Bv_{ki\pi
}^Bu_\xi ^Au_\pi ^A\right.  \nonumber \\
&&\left. +\frac 1{6\left( \Theta ^B\right) ^3}v_{ki\xi }^Bv_{ki\pi
}^Bv_{ki\eta }^Bu_\xi ^Au_\pi ^Au_\eta ^A\right\} +\cdots \text{.}
\label{DC3}
\end{eqnarray}
where 
\begin{equation}
G_k^B=\frac 1{2\pi \Theta ^B}\exp \left[ -\frac{(v_k^B)^2}{2\Theta ^B}%
\right] \text{.}  \label{DC4}
\end{equation}
The formulae for $G_k^B$ can be obtained through formal replacements in Eqs.(%
\ref{NFk1})-(\ref{F0N}): $\Theta ^A\rightarrow \Theta ^B=k_BT/m^B$, $%
F_k^A\rightarrow $ $G_k^B$.

\subsection{Models for the complete Euler equations}

LBMs for single-component Euler equation have been constructed by several
authors. [See Yan, et al\cite{PRE59454} and Kataoka, 
et al\cite{KataokaEuler}
for examples.] In this section we formulate the discrete-velocity kinetic
models A, D and E for the complete Euler equations of binary fluids.

\subsubsection{Discrete-velocity kinetic model A}

The equations for component $A$ are the same as Eqs. 
(\ref{DB3}), (\ref{DB5}) with 
\begin{eqnarray}
Q_{ki}^{AB} &=&-\frac{f_{ki}^{A(0)}}{\Theta ^A}\left\{ \mu
_D^A\left( {\bf v}_{ki}^A-{\bf u}^A\right) \cdot ({\bf u}^A-{\bf u}^B)+\mu
_T^A\left[ \frac{\left( {\bf v}_{ki}^A-{\bf u}^A\right) ^2}{2\Theta ^A}%
-1\right] (T^A-T^B)\right.  \nonumber \\
&&\left. -M^A\left[ \frac{\left( {\bf v}_{ki}^A-{\bf u}^A\right) ^2}{2\Theta
^A}-1\right] ({\bf u}^A-{\bf u}^B)^2\right\} \text{.}  \label{DA1}
\end{eqnarray}

The Chapman-Enskog analysis\cite{Chap} shows that, to recover the same Euler
equations, (\ref{continuity1}), (\ref{Ema}) and (\ref{Eena}), besides 
(\ref{rti1})-(\ref{rti9}), one more requirement 
\begin{eqnarray}
&&\sum_{ki}\frac 12m^A\left( v_k^A\right) ^2v_{ki\alpha }^Av_{ki\beta
}^Af_{ki}^{A(0)}  \nonumber \\
&=&2P^A_0 \Theta ^A\delta _{\alpha \beta }+\frac{P^A_0}2\left( u^A\right)
^2\delta _{\alpha \beta }+3P^Au_\alpha ^Au_\beta ^A+\frac 12\rho ^A\left(
u^A\right) ^2u_\alpha ^Au_\beta ^A  \label{rt12a}
\end{eqnarray}
is necessary. Correspondingly, $f_{ki}^{A(0)}$ should be Taylor expanded to
the fourth order of flow velocity, the DVM should be isotropic up to its
seventh rank tensor. Again, DVM (\ref{i2}) satisfies the need.
To satisfy (\ref{rt12a}), we require 
\begin{equation}
\sum_{ki}F_k^A\frac{\left( v_k^A\right) ^2}2v_{ki\alpha }^Av_{ki\beta
}^A=2\left( \Theta ^A\right) ^2\delta _{\alpha \beta }\text{ ,}
\label{rt13r1}
\end{equation}
\begin{equation}
\sum_{ki}F_k^A\frac{\left( v_k^A\right) ^2}2v_{ki\alpha }^Av_{ki\beta
}^Av_{ki\xi }^Av_{ki\pi }^Au_\xi ^Au_\pi ^A=3\left( \Theta ^A\right)
^3\left[ \left( u^A\right) ^2\delta _{\alpha \beta }+2u_\alpha ^Au_\beta
^A\right] \text{,}  \label{rt13r2a}
\end{equation}
\begin{eqnarray}
&&\sum_{ki}F_k^A\frac{\left( v_k^A\right) ^2}2v_{ki\alpha }^Av_{ki\beta
}^Av_{ki\xi }^Av_{ki\pi }^Av_{ki\eta }^Av_{ki\lambda }^Au_\xi ^Au_\pi
^Au_\eta ^Au_\lambda ^A  \nonumber \\
&=&12\left( \Theta ^A\right) ^4\left[ \left( u^A\right) ^4\delta _{\alpha
\beta }+4\left( u^A\right) ^2u_\alpha ^Au_\beta ^A\right] \text{.}
\label{rt13r3}
\end{eqnarray}
Finally, we have five requirements on $F_k^A$. Four are shown in Eqs. (\ref
{rei1})-(\ref{rei4}) and the fifth is 
\begin{equation}
\sum_kF_k^A\left( v_k^A\right) ^8=48\left( \Theta ^A\right) ^4\text{.}
\label{re5}
\end{equation}
To satisfy the above five requirements, five particle velocities are
sufficient. We choose a zero speed, $v_0^A=0$, and other four nonzero ones, $%
v_k^A$ ($k=1$, $2$, $3$, $4$). It is easy to find the following solution, 
\begin{eqnarray}
F_0^A &=&1-8\sum_{k=1}^4F_k^A\text{, }  \label{Fk0} \\
F_k^A &=&\frac{\Psi _k^A}{\Phi _k^A}\text{, }  \label{Fkn}
\end{eqnarray}
where 
\begin{equation}
\Psi _k^A=192\left( \Theta ^A\right) ^4-24\left( \Theta ^A\right)
^3\sum_{j=1}^3\left( v_{k+j}^A\right) ^2+4\left( \Theta ^A\right)
^2\sum_{j=1}^3\left( v_{k+j}^Av_{k+j+1}^A\right) ^2-\Theta ^A\Pi
_{j=1}^3\left( v_{k+j}^A\right) ^2  \label{Fk2}
\end{equation}
\begin{equation}
\Phi _k^A=4\left( v_k^A\right) ^2\Pi _{j=1}^3\left[ \left( v_k^A\right)
^2-\left( v_{k+j}^A\right) ^2\right] \text{, }  \label{Fk3}
\end{equation}
$k=1$,$2$,$3$,$4$, and $v_{4+j}^A=v_j^A$, ($j=1$, $2$, $3$).
For component $B$ we have similar results.

\subsubsection{Discrete-velocity kinetic model D}

The equations for component $A$ are the same as those of
model A and for component $B$ are as follows, 
\begin{equation}
Q^{BB}=-\frac 1{\tau ^{BB}}\left[ f_{ki}^B-f_{ki}^{B\left( 0\right) }\right]
-\frac 1{\tau ^{BA}}\left[ f_{ki}^B-g_{ki}^{B\left( 0\right) }\right] \text{,%
}  \label{DD1}
\end{equation}
\begin{eqnarray}
Q_{ki}^{BA} &=&-\frac{g_{ki}^{B\left( 0\right) }}{\Theta ^{Br}}\left\{ \mu
_D^{B*}\left( {\bf v}_{ki}^B-{\bf u}^A\right) \cdot ({\bf u}^A-{\bf u}%
^B)+\mu _T^{B*}\left[ \frac{\left( {\bf v}_{ki}^B-{\bf u}^A\right) ^2}{%
2\Theta ^{Br}}-1\right] \left( T^A-T^B\right) \right.  \nonumber \\
&&\left. -M^B\left[ \frac{\left( {\bf v}_{ki}^B-{\bf u}^A\right) ^2}{2\Theta
^{Br}}-1\right] ({\bf u}^B-{\bf u}^A)^2\right\} \text{.}  \label{DD2}
\end{eqnarray}

$f_{ki}^{A\left( 0\right) }$ and $f_{ki}^{B\left(0\right) }$ are
formulated in the same way as in model A. $g_{ki}^{B\left( 0\right) }$ 
is expanded in the similar way to (\ref{DC3}) but to the fourth order 
of flow velocity. The formulae for $G_k^B$ can be obtained through
formal replacements in Eqs.(\ref{Fk0})-(\ref{Fk3}): $\Theta ^A\rightarrow \Theta ^{Br}=k_BT^A/m^B$, $F_k^A\rightarrow $ $G_k^B$.

\subsubsection{Discrete-velocity kinetic model E}

Within model E 
\begin{eqnarray}
Q_{ki}^{AA} &=&-\frac 1{\tau ^{AA}}\left[ f_{ki}^A-f_{ki}^{A\left(
0\right) }\right] -\frac 1{\tau ^{AB}}\left[ f_{ki}^A-g_{ki}^{A\left(
0\right) }\right] \text{,}  \label{DE1} \\
Q_{ki}^{BB} &=&-\frac 1{\tau ^{BB}}\left[ f_{ki}^B-f_{ki}^{B\left( 0\right)
}\right] -\frac 1{\tau ^{BA}}\left[ f_{ki}^B-g_{ki}^{B\left( 0\right)
}\right] \text{,}  \label{DE2}
\end{eqnarray}
\begin{eqnarray}
Q_{ki}^{AB} &=&-\frac{g_{ki}^{A\left( 0\right) }}{\Theta ^{Ar}}%
\left\{ \mu _D^A\left( {\bf v}_{ki}^A-{\bf u}^A\right) \cdot ({\bf u}^A-{\bf %
u}^B)+\mu _T^A\left[ \frac{\left( {\bf v}_{ki}^A-{\bf u}^A\right) ^2}{%
2\Theta ^{Ar}}-1\right] (T^A-T^B)\right.  \nonumber \\
&&\left. -M^A\left[ \frac{\left( {\bf v}_{ki}^A-{\bf u}^A\right) ^2}{2\Theta
^{Ar}}-1\right] ({\bf u}^A-{\bf u}^B)^2\right\}\text{,}  \label{DE3}
\end{eqnarray}
\begin{eqnarray}
Q_{ki}^{BA} &=&-\frac{g_{ki}^{B\left( 0\right) }}{\Theta ^B}\left\{ \mu
_D^{B*}\left( {\bf v}_{ki}^B-{\bf u}^A\right) \cdot ({\bf u}^A-{\bf u}%
^B)+\mu _T^{B*}\left[ \frac{\left( {\bf v}_{ki}^B-{\bf u}^A\right) ^2}{%
2\Theta ^B}-1\right] \left( T^A-T^B\right) \right.  \nonumber \\
&&\left. -M^B\left[ \frac{\left( {\bf v}_{ki}^B-{\bf u}^A\right) ^2}{2\Theta
^B}-1\right] ({\bf u}^B-{\bf u}^A)^2\right\} \text{.}  \label{DE4}
\end{eqnarray}

$f_{ki}^{A\left( 0\right) }$, $f_{ki}^{B\left( 0\right) }$ are formulated in
the same way as in model A. The formulations of $%
g_{ki}^{A\left( 0\right) }$ and $g_{ki}^{B\left( 0\right) }$ are similar to
those in model D. The requirements on $g_{ki}^{B(0)}$ can
be obtained from those of model D by using formal
replacements: 
${\bf u}^B\rightarrow {\bf u}^A$, 
$\Theta ^{Br}\rightarrow\Theta ^B$. 
Then the requirements on $g_{ki}^{A(0)}$ can be obtained from
those on $g_{ki}^{B(0)}$ by using formal replacements: 
${\bf v}^B\rightarrow {\bf v}^A$, 
$\Theta ^B\rightarrow \Theta ^{Ar}=k_BT^B/m^A$, 
$g_{ki}^{B(0)}\rightarrow g_{ki}^{A(0)}$.

\subsection{Finite-difference schemes, spurious viscosities and diffusivities
}

The time evolution of a discrete-velocity kinetic model can be solved
numerically by using appropriate finite-difference schemes. There are
various options for calculating the time derivative and the advection 
term\cite{CEJP2382,VictorIJMPC,JCP184422}. 

In a practical simulation the real evolution equation of $f_{ki}^A$ is not
Eq. (\ref{DB1}) but
\begin{equation}
\frac{\partial f_{ki}^A}{\partial t}+\theta \frac{\partial ^2}{\partial t^2}%
f_{ki}^A+\left( v_{ki\alpha }^A\frac{\partial f_{ki}^A}{\partial \alpha }%
+\psi v_{ki\alpha }^A\frac{\partial ^2f_{ki}^A}{\partial \alpha ^2}\right)
=\left[ Q_{ki}^{AA}+Q_{ki}^{AB}+{\bf a}^A\cdot \frac{\left( {\bf v}_{ki}^A-%
{\bf u}^A\right) }{\Theta ^A}f_{ki}^{A\left( 0\right) }\right] \text{,}
\label{fde1}
\end{equation}
where smaller terms in the second and higher orders of $\Delta t$ or $\Delta
\alpha $ have been neglected; the factors $\theta $ and $\psi $ can be
specified for various FD schemes. The extra terms in $\theta $ and $\psi $
contribute to the spurious viscosities and diffusivities in the simulation
results. To check the spurious viscosities and diffusivities, one needs do
again the Chapman-Enskog analysis to Eq. (\ref{fde1}) and compare the
hydrodynamic equations with those of the continuous models. 
The recovered mass and momentum equations from (\ref{fde1}) are 
\begin{equation}
\frac{\partial \rho ^A}{\partial t}+\frac \partial {\partial r_\alpha }%
\left( \rho ^Au_\alpha ^A\right) =-\frac{\partial j_\alpha ^{AB,S}}{\partial
r_\alpha }\text{,}  \label{dmassA1}
\end{equation}
\begin{equation}
\frac{\partial \rho ^A}{\partial t}+\frac \partial {\partial r_\alpha }%
\left( \rho ^Au_\alpha \right) =-\frac \partial {\partial r_\alpha }\left[
j_\alpha ^{AB}+j_\alpha ^{AB,S}\right] \text{,}  \label{dmassA2}
\end{equation}
and 
\begin{eqnarray}
&&\frac \partial {\partial t}\left( \rho ^Au_\alpha ^A\right) {\bf +}\frac %
\partial {\partial r_\beta }\left( \rho ^Au_\alpha ^Au_\beta ^A\right) +%
\frac{\partial P_0^A}{\partial r_\alpha }-\rho ^Aa_\alpha ^A+\frac{j_\alpha
^{AB}}{\tau ^{AB}}  \nonumber \\
&&-\left[ \frac \partial {\partial r_\beta }\left( 1-\theta \frac \partial {%
\partial t}+\psi \frac \partial {\partial r_\beta }\right) \right] \Pi
_{\alpha \beta }^A  \nonumber \\
&=&\left( \theta \frac \partial {\partial t}-\psi \frac \partial {\partial
r_\beta }\right) \left[ \frac \partial {\partial r_\beta }\left( \rho
^Au_\alpha ^Au_\beta ^A\right) +\frac{\partial P_0^A}{\partial r_\beta }%
\delta _{\alpha \beta }\right] +\theta \frac \partial {\partial t}\left( 
\frac{j_\alpha ^{AB}}{\tau ^{AB}}-\rho ^Aa_\alpha ^A\right)  \label{dmomA}
\end{eqnarray}
where 
\begin{equation}
j_\alpha ^{AB,S}=-\left( \theta \frac \partial {\partial t}-\psi \frac %
\partial {\partial r_\alpha }\right) \left( \rho ^Au_\alpha ^A\right)
\label{jABs}
\end{equation}
is the spurious diffusion flux density, $\partial /\partial r_\beta
\left( \partial /\partial r_\beta \right) =\partial ^2/\partial r_\beta ^2$.
The spurious diffusivity and viscosity are coupled in the real momentum
equation (\ref{dmomA}). The real momentum equations for component $B$
 and the real energy equations can be
considered in a similar way.

Which FD scheme to use depends on the question under consideration. Since
the higher-order schemes for time derivative require more memory, the
forward Euler scheme is generally used. In binary systems concentration 
gradients drive the diffusion behavior. For systems  with large density 
gradients, 
the space centered scheme is less stable and
the wiggle phenomena of the second-order upwind, the
Lax-Wendroff and the Beam-Warming schemes introduces
unphysical oscillations of fluid
densities\cite{CEJP2382,VictorIJMPC,JCP184422}. Therefore, for such a 
system, the first-order upwind scheme 
\begin{equation}
\frac{\partial f_{ki}^A}{\partial \alpha }=\left\{ 
\begin{array}{ll}
\frac{f_{ki,I}^A-f_{ki,I-1}^A}{\Delta \alpha } & \text{if }v_{ki\alpha
}^A\geq 0 \\ 
\frac{f_{ki,I}^A-f_{ki,I+1}^A}{-\Delta \alpha } & \text{if }v_{ki\alpha }^A<0
\end{array}
\right. \text{,}  \label{fds1}
\end{equation}
is generally preferred, where the third subscripts $I-1$, $I$, $I+1$ in Eq. (%
\ref{fds1}) indicate consecutive mesh nodes in the $\alpha $ direction and $%
\Delta \alpha $ is the space step. In such a FDLBM scheme, $\theta =\Delta
t/2$, $\psi =-\Delta \alpha /2$ if  $v_{ki\alpha }^A\geq 0$ and $\psi
=\Delta \alpha /2$ if  $v_{ki\alpha }^A<0$.

It should be noted that besides the FD schemes and truncation errors of
the machine, the numerical errors from
the DVMs also contribute to spurious diffusivities and/or viscosities. The
smaller the hydrodynamic velocity, the less this part of
contribution. Other discussions on the origin of spurious velocities and
possible remedies are referred to 
\cite{PRE6756105,CEJP2382,VictorIJMPC,JCP184422,Wagner,DR,SetaTakahashi,HeLuo,Fang}.

\section{Numerical tests}

As mentioned above, in a practical simulation the numerical errors have three resources, the
formulated DVM, the spacial discretization and the time discretization.
We first check the case where the spacial FD scheme has no contribution 
-- the uniform relaxation process where the physical quantities are only 
functions of time.
For the velocity equilibration, the five kinetic models give the
same expression, 
\begin{equation}
\frac{\partial \left( u_\alpha ^A-u_\alpha ^B\right) }{\partial t}=-\frac 1%
\rho \left( \frac{\rho ^B}{\tau ^{AB}}+\frac{\rho ^A}{\tau ^{BA}}\right)
\left( u_\alpha ^A-u_\alpha ^B\right) \text{.}  \label{Ueq}
\end{equation}
For temperature equilibration, model A gives 
\begin{eqnarray}
\frac{\partial \left( T^A-T^B\right) }{\partial t} &=&-\frac 1n\left( \frac{%
n^B}{\tau ^{AB}}+\frac{n^A}{\tau ^{BA}}\right) \left( T^A-T^B\right)  
\nonumber \\
&&+\frac{\rho ^A\rho ^B}{2k_Bn\rho }\left( \frac 1{\tau ^{BA}}-\frac 1{\tau
^{AB}}\right) \left( {\bf u}^A-{\bf u}^B\right) ^2\text{,}  \label{TeqA}
\end{eqnarray}
model D gives 
\begin{eqnarray}
\frac{\partial \left( T^A-T^B\right) }{\partial t} &=&-\frac 1n\left( \frac{%
n^B}{\tau ^{AB}}+\frac{n^A}{\tau ^{BA}}\right) \left( T^A-T^B\right)  
\nonumber \\
&&+\frac 1{2k_Bn\rho }\left[ \rho ^A\rho ^B\left( \frac 1{\tau ^{AB}}-\frac 1%
{\tau ^{BA}}\right) -\frac{nm^B}{\tau ^{BA}}\left( \rho ^A-\rho ^B\right)
\right] \left( {\bf u}^A-{\bf u}^B\right) ^2\text{,}  \label{TeqD}
\end{eqnarray}
and model E gives 
\begin{eqnarray}
\frac{\partial \left( T^A-T^B\right) }{\partial t} &=&-\frac 1n\left( \frac{%
n^B}{\tau ^{AB}}+\frac{n^A}{\tau ^{BA}}\right) \left( T^A-T^B\right)  
\nonumber \\
&&+\frac 1{2k_Bn\rho }\left[ \rho ^A\rho ^B\left( \frac 1{\tau ^{AB}}-\frac 1%
{\tau ^{BA}}\right) -\frac{nm^B}{\tau ^{BA}}\left( \rho ^A-\rho ^B\right)
\right] \left( {\bf u}^A-{\bf u}^B\right) ^2\text{.}  \label{TeqE}
\end{eqnarray}
Numerical examples are shown in Fig.1. In Fig. 1 (a) we show two cases where 
$\rho ^A=\rho ^B$; in the isothermal case kinetic models A and B are
applied, while in the case of $T^A\approx T^B$ only kinetic model A is 
applied.
Fig. 1 (b) shows cases where $\rho ^A\gg \rho ^B$ so that models A
and B do not work and one has to resort on models C, D and E. For the
velocity equilibration procedure, under the accuracy of the
calculations, models A and B give the same results,
models C, D and E give the same results. All
the numerical results in (a) and (b) agree well with the theoretical
ones. 

Secondly, we check a case where the advection terms make effects and 
 viscosities exist. We use the two-fluid FDLBMs A and B to investigate the
isothermal Couette flow for single-component fluid. The two walls, locating
at  $y=\pm D/2$, start to move horizontally with velocities $\pm U$ at $t=0$,
where $D$ is the distance between the two walls. 
The simulation results of the velocity profiles agree well with the
following theoretical one, 
\begin{equation}
u=\gamma y-\sum_{j=1}^{\infty}
(-1)^{j+1}\frac{\gamma D}{j\pi }\exp (-\frac{4j^2\pi ^2\eta 
}{\rho D^2}t)\sin (\frac{2j\pi }Dy)\text{,}  \label{uy}
\end{equation}
where $u$ is the horizontal velocity, $\gamma =2U/D$ the imposed the
shear rate, $j$  an integer . 
( An example is referred to Fig. 2.)

Thirdly, we investigate the diffusion behavior in a one-dimensional
 system. To
make valid the relation (\ref{Difa}) and make less the numerical errors from
the spacial FD scheme, we assume that (i) the two components have equal
particle masses $1$, (ii) the initial hydrodynamic velocities of the two
components are zero, (iii) the system is isothermal with temperature $T=1$,
(iv) the initial density profiles of the two components are
\begin{equation}
\rho ^A=\left\{ 
\begin{array}{ll}
1-\Delta \rho  & \text{if }x\leq 0 \\ 
1+\Delta \rho  & \text{if }x>0
\end{array}
\right. \text{, }\rho ^B=\left\{ 
\begin{array}{ll}
1+\Delta \rho  & \text{if }x\leq 0 \\ 
1-\Delta \rho  & \text{if }x>0
\end{array}
\right. \text{,}  \label{rhoi}
\end{equation}
where $0\leq \Delta \rho \leq 1$, and (v) the viscosities of the two
components are small enough. Thus, the barycentric velocity field of this
system is globally zero, ${\cal D}^A=\tau ^{AB}$, ${\cal D}^B=\tau ^{BA}$,
and the evolution of the density profiles follows 
\begin{equation}
\rho ^A=1+\Delta \rho 
\mathop{\rm erf}%
\left( \frac x{\sqrt{4{\cal D}^At}}\right) \text{,}\rho ^B=1-\Delta \rho 
\mathop{\rm erf}%
\left( \frac x{\sqrt{4{\cal D}^Bt}}\right) \text{.}  \label{rhoAB}
\end{equation}
To make the numerical tests practical, when choose parameters for
simulations, the following points should be considered: (i) The accuracy of
the forward Euler scheme is in the order of $\Delta t$ and that of the
upwind scheme (\ref{fds1}) is in the order of $\Delta x$; (ii) If the
physical values of ${\cal D}^A$ and ${\cal D}^B$ are too small, they may be
submerged by the numerical diffusivities. Numerical tests show that LBMs A
and B can recover density profiles which agree well with Eq. (\ref{rhoAB}).
An example is shown in Fig.3. A set of density profiles for the component A
are shown in (a). To help evaluate the numerical errors from the DVM, the
corresponding profiles of diffusion velocity $u^A$ are shown in (b). The
diffusion velocity $u^A$ has its maximum value at $x=0$. Its magnitude
decreases with time. For the earliest time ($t=0.01$) shown in this
figure, 
$\left| u^A\right| \approx 0.028$.  The numerical errors for 
$f_{ki}^{A\left(0\right) }$ are in the order of $\left( u^A\right) ^4$ 
for LBM B and in the order of $\left( u^A\right) ^5$ for LBM A.

\section{Conclusions and remarks}

Sirovich's original two-fluid BGK kinetic theory works for symmetric systems
where the two components have approximately the same total masses and local
temperatures. This theory is clarified and generalized to describe both
symmetric and asymmetric systems. Corresponding to different situations five
kinetic models are formulated. Based on an octagonal discrete velocity model
the five models are discretized. The discrete-velocity kinetic models and the continuous ones
are required to recover the same Euler and/or Navier-Stokes equations. A
discrete-velocity kinetic model and an appropriate finite-difference scheme compose a
FDLBM. The formulated kinetic models work also for binary mixtures with
disparate particle-mass components. Which model to use depends on the mean
temperatures and the mean mass densities of the two components. 

In the present two-fluid treatment, the relaxation times of the
cross-collisions contribute to both the viscous and diffusive effects.
The interfacial tension is another aspect of thermodynamic
interaction between component fluids. 
Investigating the interfacial tension is crucial in the industrial context
for controlling the size and phase stability of mechanically dispersed
droplets and other transient structures formed in the course of phase
separation. For immiscible fluids, one way to
introduce the interfacial tension is through modifying the pressure
tensors\cite{EPL32463} by
taking into account the the
interparticle interactions. One possibility of incorporating the
interparticle interaction is through modifying the force terms in
Boltzmann equations\cite{VictorPRE}.
In such a case, the force terms in the BGK kinetic models are responsible
for the phase separation and interfacial tension. The acceleration ${\bf a}^A
$ is determined by the interparticle interactions and the external field.
The determination of the specific form of ${\bf a}^A$ depends on the system
under consideration. An interesting point is that the incorporation of the
force term in the Boltzmann equation makes no additional requirement on the
formulation process of the FDLBM. So the specific forces can be directly
considered under the same frame. 
A different attempt to introduce the
interfacial tension is to start from the Enskog equations for dense
gases\cite{PRE6835302}.

\acknowledgments
The author thanks Prof. G. Gonnella for guiding him into the LBM field and
Profs. H. Hayakawa, V. Sofonea, M. Watari, and S. Succi for helpful
discussions. The valuable comments and suggestions of the anonymous referee
is gratefully acknowledged. This work is partially supported by
Grant-in-Aids for Scientific Research (Grant No. 15540393) and for the 21-th
Century COE ``Center for Diversity and Universality in Physics'' from the
Ministry of Education, Culture and Sports, Science and Technology (MEXT) of
Japan.

\newpage
\begin{figure}
\epsfig{file=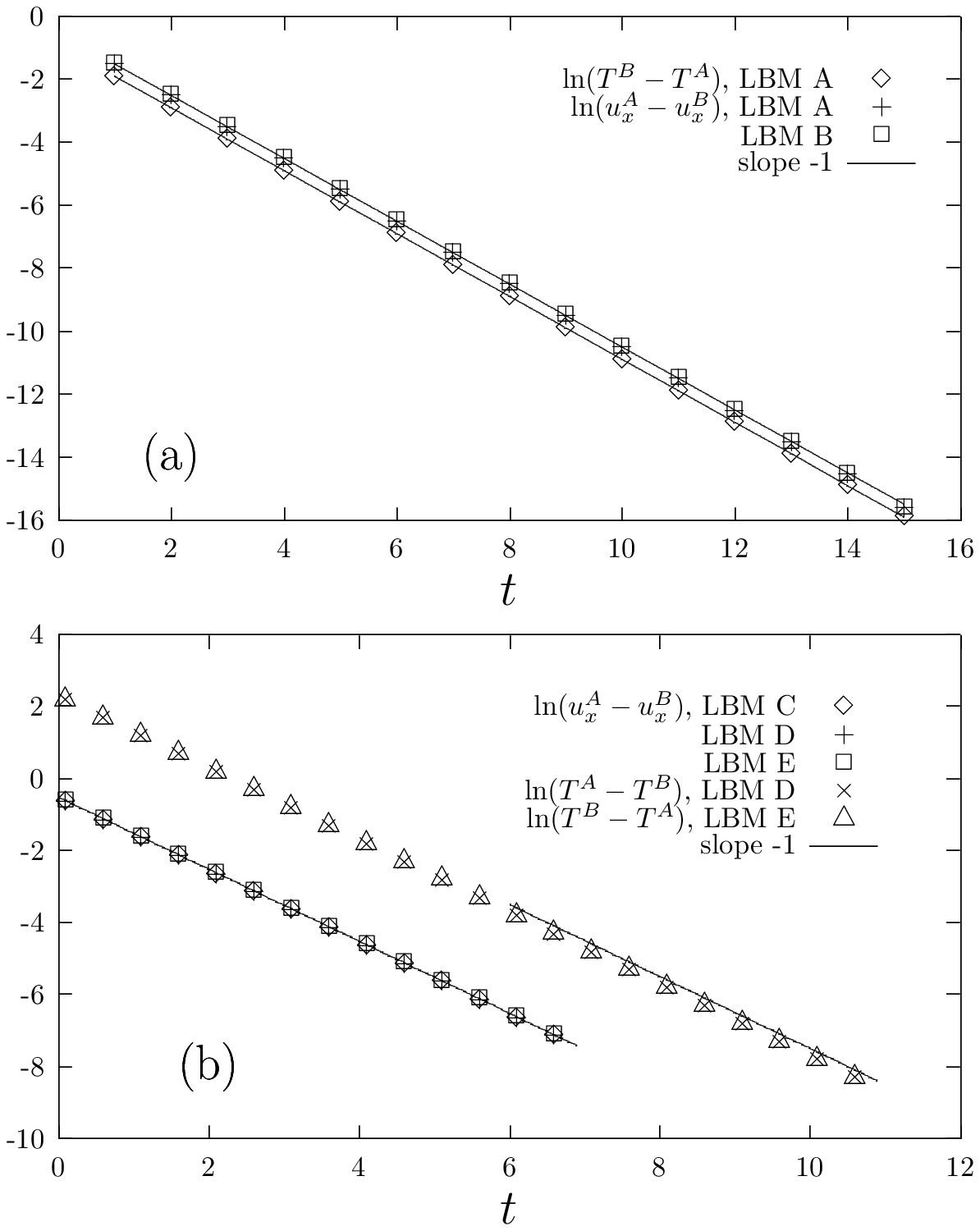,
bbllx=98 pt,bblly= 220 pt,
bburx= 476 pt,bbury= 671 pt,
width=8.5cm,clip=}
\caption{Velocity and temperature equalibrations in uniform 
relaxation processes. The common parameters used in (a) are 
$n^A=1$,$n^B=2$, $m^A=2$, $m^B=1$, 
$\tau^{AA}=\tau^{BB}=\tau^{AB}=\tau^{BA}=1$. Additionally,  
$u^{A(0)}_x=-u^{B(0)}_x=0.3$, $u^{A(0)}_y=u^{B(0)}_y=0$
 and $T=1$ for the isothermal case;  $T^A=1.2$, $T^B=0.8$, 
${\bf u}^{A(0)}={\bf u}^{B(0)}={\bf 0}$ for the thermal case.
The common parameters used in (b) are $n^A=n^B=1$, $m^A=100$, $m^B=1$,
$\tau^{AA}=\tau^{BB}=\tau^{AB}=\tau^{BA}=1$, 
$u^{A(0)}_x=-u^{B(0)}_x=0.3$,
$u^{A(0)}_y=u^{B(0)}_y=0$.
Additionally, $T=1$ for the isothermal case; 
$T^{A(0)}=10$, $T^{B(0)}=0.1$ for one and $T^{A(0)}=0.1$, $T^{B(0)}=10$
 for the other thermal cases. 
The second superscript ``(0)'' denotes the initial values. Solid lines 
in the figure possess corresponding theoretical slopes.
}
\end{figure}

\newpage

\begin{figure}
\epsfig{file=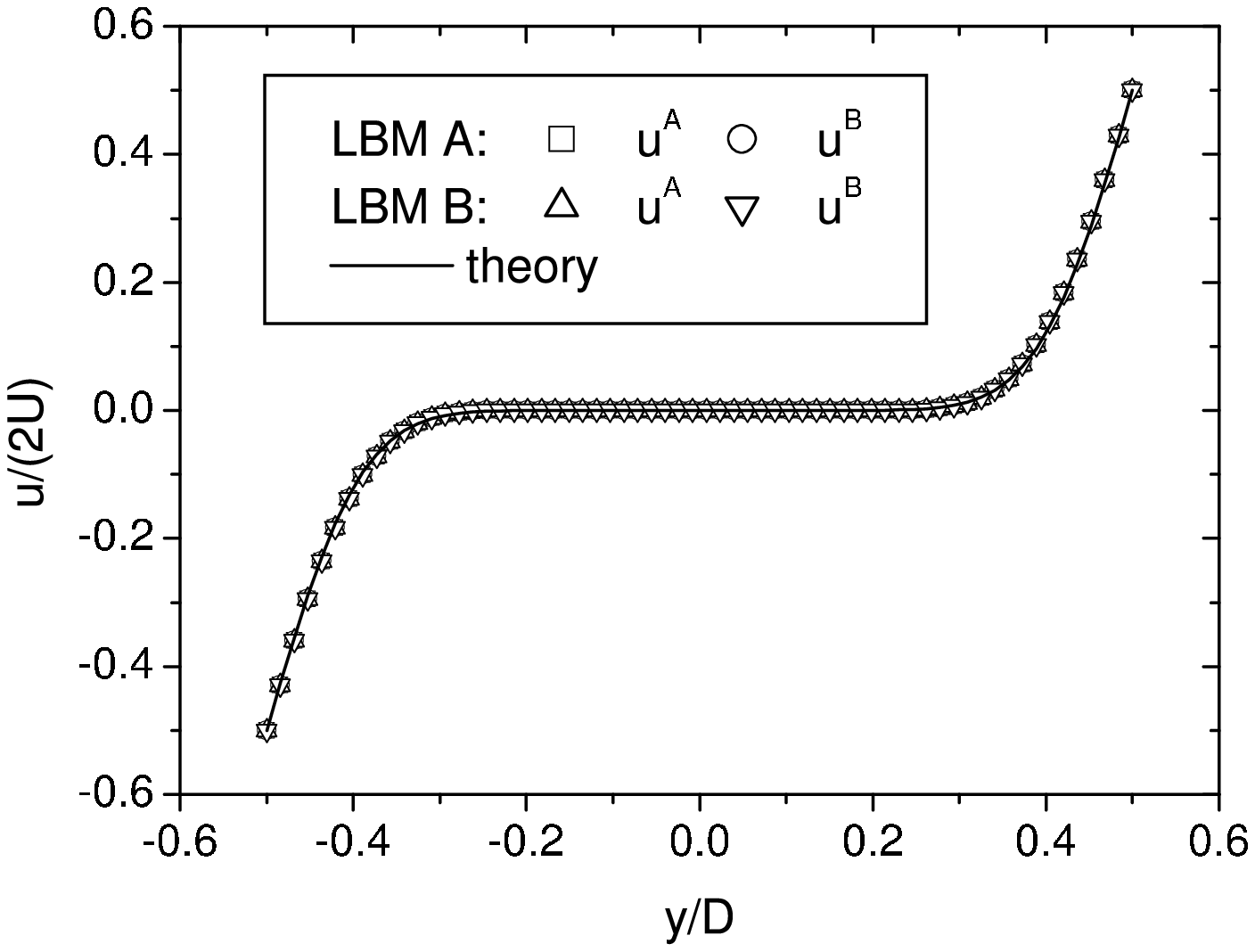,
bbllx=70 pt,bblly= 399 pt,
bburx= 475 pt,bbury= 715 pt,
width=8.5cm,clip=}
\caption{Horizontal velocity profiles along a vertical line 
for the two components at time $t=2.9$. 
The symbols denote simulation results from LBMs A and B. The solid line
shows the analytical result.
Parameters used in the simulations are 
$m^A=m^B=1$, $T^A=T^B=T_{\text{up}}=T_{\text{low}}=T=1$, $n^A=n^B=1$,
$\gamma =0.001$.
Parameters used in Eq. (\ref{uy}) are $\eta =0.05$, $\rho =1$. 
}
\end{figure}

\newpage

\begin{figure}
\epsfig{file=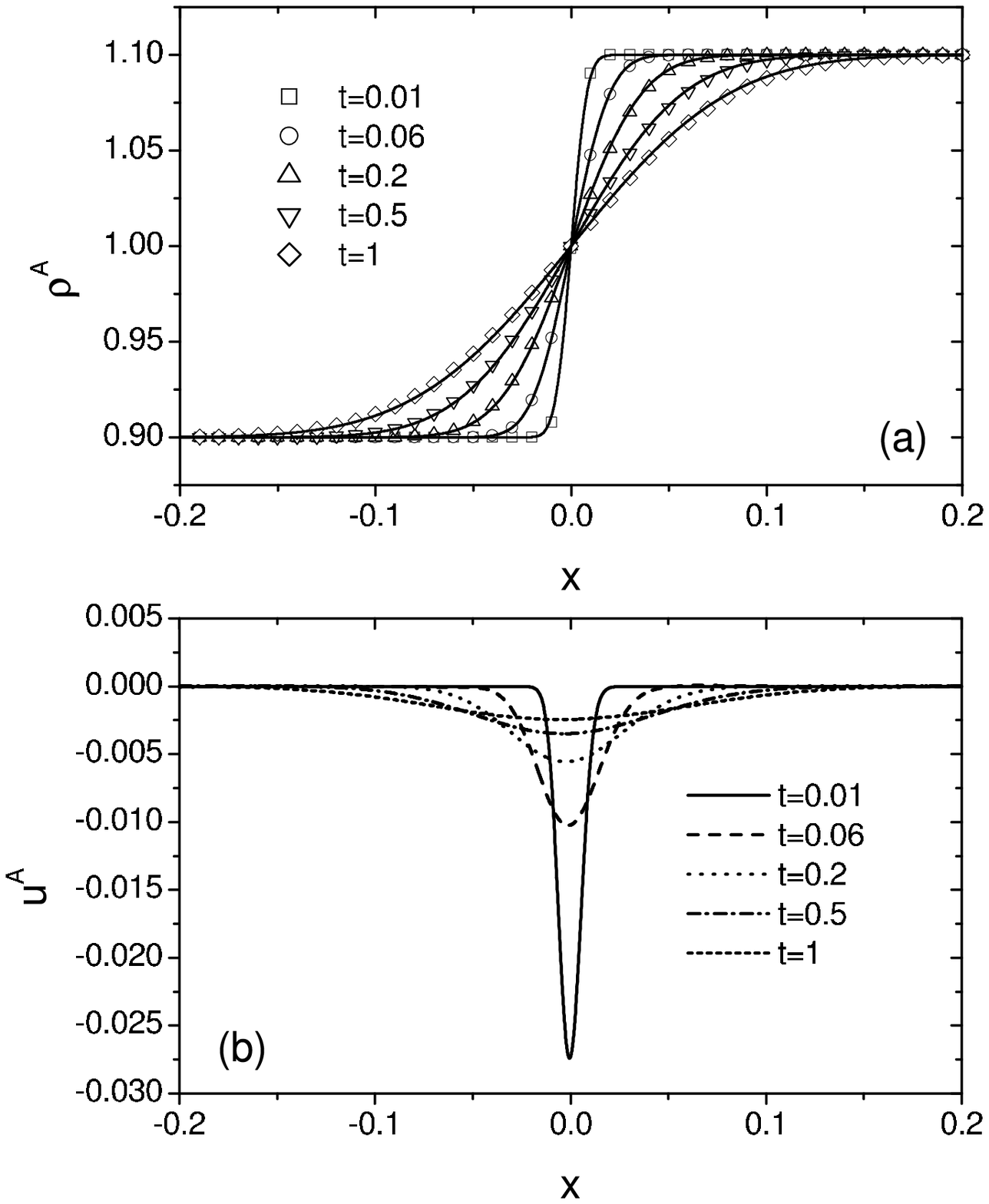,
bbllx=78 pt,bblly= 198 pt,
bburx= 480 pt,bbury= 685 pt,
width=8.5cm,clip=}
\caption{LBM simulation on a diffusion process. A set of density
 profiles of component $A$ are shown in (a) and the corresponding
 diffusion velocities are shown in (b). The initial density
 profiles of the two components follow Eq.(\ref{rhoi}) with 
$\Delta \rho=0.1$. All the
 relaxation times are taken to be $2\times 10^{-3}$. The integration 
steps are $\Delta x = \Delta y = 10^{-4}$ and $\Delta t=10^{-5}$. The
 density profiles from the simulation (symbols) agree well with the 
theoretical ones (lines). The simulation tool for this figure is LBM A. 
}

\end{figure}


\begin{thebibliography}{99}
\bibitem{foot1}The general BGK kinetic 
model of the single-component fluid can be regarded as a special case
 where $m^A = m^B$, $n^A = n^B$, ${\bf u}^A = {\bf u}^B$, $T^A = T^B$ and
 consequently the cross-collision and self-collision are the same. In
 the Chapman-Enskog analysis we do not assume that the cross-collision 
terms  are the first order in the Knudsen number.
\bibitem{foot2} At the Euler level, $f^A = f^{A(0)}$ and $f^B = f^{B(0)}$;
 but it is not required that $f^{AB(0)}=f^{A(0)}$ or $f^{BA(0)}=f^{B(0)}$,
 which is responsible for the momentum and heat exchanges between the
 two components. In the case of single-component fluid\cite{foot1}, the
 terms in momentum and heat exchanges disappear.
\end{thebibliography}
\end{document}